\documentclass[a4paper]{amsart}
\usepackage{amsmath,amsthm,amssymb}
\usepackage{epic,eepic}
\usepackage{epsfig,cite,indentfirst,oldgerm}

\begin{document}

\title[DW partition functions and KP]
{Domain wall partition functions and KP}

\author{O Foda, M Wheeler and M Zuparic}

\address{Department of Mathematics and Statistics,
         University of Melbourne,
         Parkville, Victoria 3010, Australia.}
\email{foda, mwheeler, mzup@ms.unimelb.edu.au}


\newcommand{\field}[1]{\mathbb{#1}}
\newcommand{\C}{\field{C}}
\newcommand{\N}{\field{N}}
\newcommand{\Z}{\field{Z}}
\newcommand{\R}{\field{R}}

\begin{abstract}
We observe that the partition function of the six vertex model 
on a finite square lattice with domain wall boundary conditions 
is (a restriction of) a KP $\tau$ function and express it as an 
expectation value of charged free fermions (up to an overall 
normalization). 
\end{abstract}

\maketitle

\newtheorem{ca}{Figure}
\newtheorem{corollary}{Corollary}
\newtheorem{definition}{Definition}
\newtheorem{example}{Example}
\newtheorem{lemma}{Lemma}
\newtheorem{notation}{Notation}
\newtheorem{proposition}{Proposition}
\newtheorem{remark}{Remark}
\newtheorem{condition}{Condition}
\newtheorem{korepin}{ }
\newtheorem{theorem}{Theorem}

\begin{center}
{\it To Professor T Miwa on his 60th birthday.}
\end{center}

\def\ll{\left\lgroup}
\def\rr{\right\rgroup}

\newcommand{\Proof}{\medskip\noindent {\it Proof: }}
\def\no{\nonumber}
\def\ni{\noindent}
\def\proofend{\ensuremath{\square}}
\def\pr{'}
\def\beqa{\begin{eqnarray}}
\def\eeqa{\end{eqnarray}}
\def\ba{\begin{array}}
\def\ea{\end{array}}
\def\gl{\begin{swabfamily}gl\end{swabfamily}}
\def\psis{\psi^{*}}
\def\Psis{\Psi^{*}}
\def\union{\mathop{\bigcup}}
\def\vac{|\mbox{vac}\rangle}
\def\cav{\langle\mbox{vac}|}
\def\dprod{\mathop{\prod{\mkern-29.5mu}{\mathbf\longleftarrow}}}
\def\rprod{\mathop{\prod{\mkern-28.0mu}{\mathbf\longrightarrow}}}
\def\r{\rangle}
\def\l{\langle}
\def\a{\alpha}
\def\b{\beta}
\def\hb{\hat\beta}
\def\d{\delta}
\def\g{\gamma}
\def\e{\epsilon}
\def\tg{\operatorname{tg}}
\def\ctg{\operatorname{ctg}}
 \def\sh{\operatorname{sh}}
 \def\ch{\operatorname{ch}}
\def\cth{\operatorname{cth}}
 \def\th{\operatorname{th}}
\def\eps{\varepsilon}
 \def\la{\lambda}
\def\tla{\tilde{\lambda}}
\def\Gh{\widehat{\Gamma}}
\def\tmu{\tilde{\mu}}
\def\s{\sigma}
\def\sul{\sum\limits}
\def\pl{\prod\limits}
\def\lt({\left(}
\def\rt){\right)}
\def\const{{\rm const}}
\def\argum{\{\mu_j\},\{\la_k\}}
\def\umarg{\{\la_k\},\{\mu_j\}}
\def\prodmu #1{\prod\limits_{j #1 k} \sinh(\mu_k-\mu_j)}
\def\prodla #1{\prod\limits_{j #1 k} \sinh(\lambda_k-\lambda_j)}
\newcommand{\bl}[1]{\makebox[#1em]{}}
\def\tr{\operatorname{tr}}
\def\Res{\operatorname{Res}}
\def\det{\operatorname{det}}
\def\ivac{\langle\Omega|}
\def\fvac{|\Omega\rangle}
\def\fdirac{|0\rangle}
\def\idirac{\langle0|}
\def\psis{\psi^{*}}
\def\Psis{\Psi^{*}}
\def\lprod{\mathop{\prod{\mkern-29.5mu}{\mathbf\longleftarrow}}}
\def\rprod{\mathop{\prod{\mkern-28.0mu}{\mathbf\longrightarrow}}}
\def\complex{\mathbb{C}}
\def\integer{\mathbb{Z}}

\newcommand{\boldN}{\boldsymbol{N}}
\newcommand{\bra}[1]{\langle\,#1\,|}
\newcommand{\ket}[1]{|\,#1\,\rangle}
\newcommand{\bracket}[1]{\langle\,#1\,\rangle}
\newcommand{\infinity}{\infty}
\newcommand{\ttop}[2]{\genfrac{}{}{0pt}{}{#1}{#2}}
\renewcommand{\labelenumi}{\S\theenumi.}

\let\up=\uparrow
\let\down=\downarrow
\let\tend=\rightarrow
\hyphenation{boson-ic
             ferm-ion-ic
             para-ferm-ion-ic
             two-dim-ension-al
             two-dim-ension-al
             rep-resent-ative
             par-tition
             anti-comm-uta-tion
	     var-i-ables
	     }

\newtheorem{Theorem}{Theorem}[section]
\newtheorem{Corollary}[Theorem]{Corollary}
\newtheorem{Proposition}[Theorem]{Proposition}
\newtheorem{Conjecture}[Theorem]{Conjecture}
\newtheorem{Lemma}[Theorem]{Lemma}
\newtheorem{Example}[Theorem]{Example}
\newtheorem{Note}[Theorem]{Note}
\newtheorem{Definition}[Theorem]{Definition}

\renewcommand{\mod}{\textup{mod}\,}
\newcommand{\wt}{\text{wt}\,}

\newcommand{\T}{{\mathcal T}}
\newcommand{\U}{{\mathcal U}}
\newcommand{\tT}{\tilde{\mathcal T}}
\newcommand{\tU}{\widetilde{\mathcal U}}
\newcommand{\Y}{{\mathcal Y}}
\newcommand{\B}{{\mathcal B}}
\newcommand{\D}{{\mathcal D}}
\newcommand{\M}{{\mathcal M}}
\renewcommand{\P}{{\mathcal P}}

\hyphenation{And-rews
             Gor-don
             boson-ic
             ferm-ion-ic
             para-ferm-ion-ic
             two-dim-ension-al
             two-dim-ension-al}

\setcounter{section}{-1}

\section{Introduction}\label{introduction}

In \cite{korepin}, Korepin introduced domain wall boundary conditions 
for the six vertex model on an $N$$\times$$N$ square lattice, and 
obtained a set of conditions that determine the corresponding 
partition function $Z_N$. In \cite{izergin}, Izergin proposed 
a determinant expression for $Z_{N}$, and showed that it satisfies 
Korepin's conditions. In \cite{lascoux}, Lascoux rewrote $Z_{N}$ 
as a determinant of a product of two non-square matrices. In this 
note we observe that, given Lascoux's form, $Z_N$ is (a restriction 
of) a KP $\tau$ function (for all values of the crossing parameter) 
and rewrite it as an expectation value of charged free fermions. 

In section {\bf \ref{DWPF}}, we recall basic definitions related to 
the six vertex model and domain wall boundary conditions, including 
Korepin's conditions, and in section {\bf \ref{determinant}}, we 
recall Izergin's determinant expression for $Z_N$, followed by 
Lascoux's expression, and observe that the latter is (a restriction 
of) a KP $\tau$ function. In section {\bf \ref{ferm}}, we propose an 
expectation value of KP charged free fermions $F_N^{\textit{free}}$ 
that by construction is a KP $\tau$ function, and show that (under 
suitable restrictions of the KP time variables) $F_N^{\textit{free}}$ 
becomes equal to $Z_N$ (up to an overall normalization).
In section {\bf \ref{discussion}}, we include a number of remarks.

\section{Domain wall partition functions}
\label{DWPF}
In this section, we list a number of basic definitions 
related to the six vertex model, domain wall boundary 
conditions, {\it etc}. For details, see 
\cite{baxter-book, korepin-book}.
\subsection{Oriented lattice lines and rapidity variables}
Consider a square lattice with $N$ horizontal lines (rows)
and $N$ vertical lines (columns) that intersect at $N^2$ 
points, and assign the $i$-th horizontal line an orientation 
from left to right and a rapidity $x_i$, and the $j$-th vertical 
line an orientation from bottom to top and a rapidity $y_j$,
as in Figure {\bf \ref{lattice}}.

\subsection{Arrows, weights and a crossing parameter}
Assign each line segment an arrow that can point in either
direction, and define the vertex $v_{ij}$ as the intersection 
point of the $i$-th horizontal line and the $j$-th vertical 
line, the four line segments attached to this intersection 
point and the arrows on these segments. Assign $v_{ij}$ 
a weight $w_{ij}$ that depends on the orientations of its 
arrows, the rapidities $x_i$ and $y_j$ and a {\it crossing 
parameter} $\mu$ that is the same for all vertices.

\subsection{The six vertex model}
Since an arrow on a line segment can point in either direction, 
there are $2^4 = 16$ possible (types of) vertices. In the
six-vertex model, the weights of all vertices, except six, are 
zero \cite{baxter-book}. The six vertices with non-zero weights 
form three pairs of equal-weight vertices. They are shown in Figure 
{\bf \ref{vertices}}.

%FIG-01
%
\begin{center}
\begin{minipage}{4.2in}
\setlength{\unitlength}{0.0010cm}
\begin{picture}(5000, 6000)(-1500, 0)
% negative x shifts the figure to the right
% negative y shifts the figure up
\thicklines
\path(2400,5400)(2400,1800)
\path(3000,5400)(3000,1800)
\path(3600,5400)(3600,1800)
\path(4200,5400)(4200,1800)
\path(4800,5400)(4800,1800)
\path(1800,4800)(5400,4800)
\path(1800,4200)(5400,4200)
\path(1800,3600)(5400,3600)
\path(1800,3000)(5400,3000)
\path(1800,2400)(5400,2400)
\path(0600,4254)(1200,4254)
\whiten\path(840,4164)(1200,4254)(840,4344)(840,4164)
\path(600,3654)(1200,3654)
\whiten\path(840,3564)(1200,3654)(840,3744)(840,3564)
\path(600,4854)(1200,4854)
\whiten\path(840,4764)(1200,4854)(840,4944)(840,4764)
\path(600,3054)(1200,3054)
\whiten\path(840,2964)(1200,3054)(840,3144)(840,2964)
\path(600,2454)(1200,2454)
\whiten\path(840,2364)(1200,2454)(840,2544)(840,2364)
\path(2400,654)(2400,1254)
\whiten\path(2490,894)(2400,1254)(2310,894)(2490,894)
\path(3000,654)(3000,1254)
\whiten\path(3090,894)(3000,1254)(2910,894)(3090,894)
\path(3600,654)(3600,1254)
\whiten\path(3690,894)(3600,1254)(3510,894)(3690,894)
\path(4200,654)(4200,1254)
\whiten\path(4290,894)(4200,1254)(4110,894)(4290,894)
\path(4800,654)(4800,1254)
\whiten\path(4890,894)(4800,1254)(4710,894)(4890,894)
\put(0050,4854){$x_1$}
\put(0050,4254){$x_2$}
\put(0050,2454){$x_N$}
\put(2300,0250){$y_1$}
\put(2900,0250){$y_2$}
\put(4700,0250){$y_N$}
\end{picture}
\begin{ca}
\label{lattice}
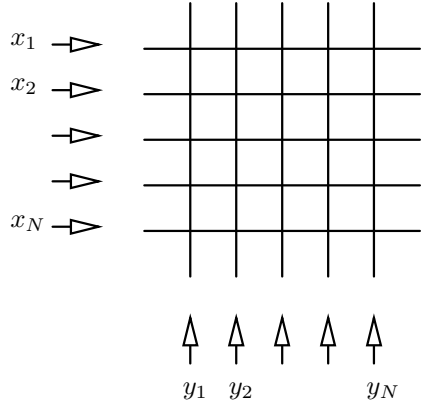
A square lattice with oriented lines and rapidity
variables. The line orientations are indicated by
white arrows. 
\end{ca}
\end{minipage}
\end{center}

%FIG-02
%
\begin{center}
\begin{minipage}{4.7in}
\setlength{\unitlength}{0.0008cm}
\begin{picture}(10000,07500)(-1000, 0)
\thicklines
\blacken\path(10162,2310)(10522,2400)(10162,2490)(10162,2310)
\blacken\path(1282,2490)(0922,2400)(1282,2310)(1282,2490)
\blacken\path(1462,5610)(1822,5700)(1462,5790)(1462,5610)
\blacken\path(1732,1860)(1822,1500)(1912,1860)(1732,1860)
\blacken\path(1732,2685)(1822,2325)(1912,2685)(1732,2685)
\blacken\path(1912,5340)(1822,5700)(1732,5340)(1912,5340)
\blacken\path(1912,6240)(1822,6600)(1732,6240)(1912,6240)
\blacken\path(2182,2490)(1822,2400)(2182,2310)(2182,2490)
\blacken\path(2362,5610)(2722,5700)(2362,5790)(2362,5610)
\blacken\path(5182,5790)(4822,5700)(5182,5610)(5182,5790)
\blacken\path(5362,2310)(5722,2400)(5362,2490)(5362,2310)
\blacken\path(5632,1860)(5722,1500)(5812,1860)(5632,1860)
\blacken\path(5632,2760)(5722,2400)(5812,2760)(5632,2760)
\blacken\path(5812,5340)(5722,5700)(5632,5340)(5812,5340)
\blacken\path(5812,6240)(5722,6600)(5632,6240)(5812,6240)
\blacken\path(6082,5790)(5722,5700)(6082,5610)(6082,5790)
\blacken\path(6262,2310)(6622,2400)(6262,2490)(6262,2310)
\blacken\path(9082,2490)(8722,2400)(9082,2310)(9082,2490)
\blacken\path(9262,5610)(9622,5700)(9262,5790)(9262,5610)
\blacken\path(9532,2760)(9622,2400)(9712,2760)(9532,2760)
\blacken\path(9532,5160)(9622,4800)(9712,5160)(9532,5160)
\blacken\path(9712,2040)(9622,2400)(9532,2040)(9712,2040)
\blacken\path(9712,6240)(9622,6600)(9532,6240)(9712,6240)
\blacken\path(9982,5790)(9622,5700)(9982,5610)(9982,5790)
\path(10222,5700)(9622,5700)
\path(1222,5700)(1822,5700)
\path(1522,2400)(0922,2400)
\path(1822,1500)(1822,2100)
\path(1822,2325)(1822,2925)
\path(1822,3300)(1822,1500)
\path(1822,3900)(1822,4500)
\path(1822,5700)(1822,5100)
\path(1822,0600)(1822,1200)
\path(1822,6600)(1822,4800)
\path(0022,2400)(0622,2400)
\path(0022,5700)(0622,5700)
\path(2422,2400)(1822,2400)
\path(3922,2400)(4522,2400)
\path(3922,5700)(4522,5700)
\path(4822,2400)(6622,2400)
\path(4822,5700)(6622,5700)
\path(5122,2400)(5722,2400)
\path(5422,5700)(4822,5700)
\path(5722,1500)(5722,2100)
\path(5722,2400)(5722,3000)
\path(5722,3300)(5722,1500)
\path(5722,3900)(5722,4500)
\path(5722,5700)(5722,5100)
\path(5722,0600)(5722,1200)
\path(5722,6600)(5722,4800)
\path(5722,6600)(5722,6000)
\path(6022,2400)(6622,2400)
\path(6322,5700)(5722,5700)
\path(7822,2400)(8422,2400)
\path(7822,5700)(8422,5700)
\path(8722,2400)(10522,2400)
\path(8722,5700)(10522,5700)
\path(9022,5700)(9622,5700)
\path(0922,2400)(2722,2400)
\path(0922,5700)(2722,5700)
\path(9322,2400)(8722,2400)
\path(9622,2400)(9622,1800)
\path(9622,2400)(9622,3000)
\path(9622,3300)(9622,1500)
\path(9622,3900)(9622,4500)
\path(9622,4800)(9622,5400)
\path(9622,0600)(9622,1200)
\path(9622,6600)(9622,4800)
\path(9622,6600)(9622,6000)
\path(9922,2400)(10522,2400)
\whiten\path(1912,4140)(1822,4500)(1732,4140)(1912,4140)
\whiten\path(1912,0840)(1822,1200)(1732,0840)(1912,0840)
\whiten\path(0262,2310)(0622,2400)(0262,2490)(0262,2310)
\whiten\path(0262,5610)(0622,5700)(0262,5790)(0262,5610)
\whiten\path(4162,2310)(4522,2400)(4162,2490)(4162,2310)
\whiten\path(4162,5610)(4522,5700)(4162,5790)(4162,5610)
\whiten\path(5812,4140)(5722,4500)(5632,4140)(5812,4140)
\whiten\path(5812,0840)(5722,1200)(5632,0840)(5812,0840)
\whiten\path(8062,2310)(8422,2400)(8062,2490)(8062,2310)
\whiten\path(8062,5610)(8422,5700)(8062,5790)(8062,5610)
\whiten\path(9712,4140)(9622,4500)(9532,4140)(9712,4140)
\whiten\path(9712,0840)(9622,1200)(9532,0840)(9712,0840)
\put(1500,0000){$w^{a}_{ij}$}
\put(5500,0000){$w^{b}_{ij}$}
\put(9500,0000){$w^{c}_{ij}$}
\put(00300,5000){$x_i$}
\put(04200,5000){$x_i$}
\put(08100,5000){$x_i$}
\put(00300,1700){$x_i$}
\put(04200,1700){$x_i$}
\put(08100,1700){$x_i$}
\put(02000,4200){$y_j$}
\put(06000,4200){$y_j$}
\put(10000,4200){$y_j$}
\put(02000,0900){$y_j$}
\put(06000,0900){$y_j$}
\put(10000,0900){$y_j$}
\end{picture}
\begin{ca}
\label{vertices}
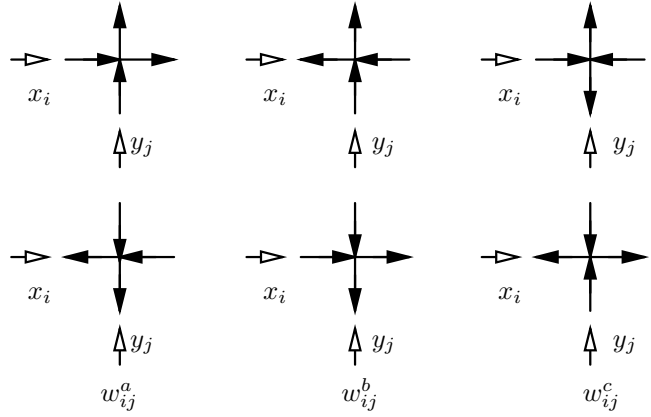
The vertices of the six vertex model. Vertices that are 
in the same column share the weight shown below that
column. 
The white arrows indicate the line orientations 
which are needed to completely specify the vertices.
\end{ca}
\end{minipage}
\end{center}
\bigskip

In the notation of Figure {\bf \ref{vertices}}, the weights 
of the six-vertex model are 
\begin{equation}
\label{weights}
w^{a}_{ij} = \sinh (-x_i + y_j + \mu),
\quad
w^{b}_{ij} = \sinh (-x_i + y_j),
\quad
w^{c}_{ij} = \sinh \mu
\end{equation}

\noindent that satisfy the Yang-Baxter equations 
\cite{baxter-book}.

\subsection{Domain wall boundary conditions} 
When all arrows on the left and right boundaries point inwards, 
and all arrows on the upper and lower boundaries point outwards, 
as in Figure {\bf \ref{dwbcfigure}}, we obtain {\it domain wall 
boundary conditions}. The inner arrows remain free. Different
inner arrow orientations lead to different {\it lattice 
configurations} that together constitute the statistical 
states of the model. 

\subsection{Domain wall partition functions} 
The weight of a lattice configuration is the product of the weights 
$w_{ij}$ of its vertices $v_{ij}$. The weighted sum over all 
lattice configurations, on an $N$$\times$$N$ lattice with fixed 
domain wall boundary conditions, is a domain wall partition 
function $Z_N$. 

\begin{equation}
\label{partitionfunctionequation}
Z_N 
\equiv 
Z_{N}\!\!\!\ll {\bf x}, {\bf y}, \mu\rr :=
\sum_{
\ttop{\rm all \phantom{-} allowed}{\rm configurations}
}
\prod_{
\ttop{\rm all}{\rm vertices}
}
w_{ij}(x_i, y_j, \mu)
\end{equation}

%FIG-03
%
\begin{center}
\begin{minipage}{4.3in}
\setlength{\unitlength}{0.0008cm}
\begin{picture}(6000,6500)(-4000, 0)
% negative x shifts the figure to the right -->
% negative y shifts the figure up
\thicklines
\path(0300,4800)(5700,4800)
\path(0300,3900)(5700,3900)
\path(0300,2100)(5700,2100)
\path(0300,1200)(5700,1200)
\path(1200,5700)(1200,0300)
\path(2100,5700)(2100,0300)
\path(4800,5700)(4800,0300)
\path(1200,5025)(1200,5625)
\blacken\path(1290,5265)(1200,5625)(1110,5265)(1290,5265)
\path(2100,5025)(2100,5625)
\blacken\path(2190,5265)(2100,5625)(2010,5265)(2190,5265)
\path(3900,5025)(3900,5625)
\blacken\path(3990,5265)(3900,5625)(3810,5265)(3990,5265)
\path(4800,5100)(4800,5700)
\blacken\path(4905,5340)(4800,5700)(4695,5340)(4905,5340)
\path(1425,4800)(2025,4800)
\blacken\path(1665,4710)(2025,4800)(1665,4890)(1665,4710)
\path(3675,4800)(3075,4800)
\blacken\path(3435,4890)(3075,4800)(3435,4710)(3435,4890)
\path(4500,4800)(3900,4800)
\blacken\path(4260,4890)(3900,4800)(4260,4710)(4260,4890)
\path(5400,4800)(4800,4800)
\blacken\path(5160,4890)(4800,4800)(5160,4710)(5160,4890)
\path(525,3900)(1125,3900)
\blacken\path(0765,3810)(1125,3900)(0765,3990)(0765,3810)
\path(1800,3900)(1200,3900)
\blacken\path(1560,3990)(1200,3900)(1560,3810)(1560,3990)
\path(5400,3900)(4800,3900)
\blacken\path(5160,3990)(4800,3900)(5160,3810)(5160,3990)
\path(600,3000)(1200,3000)
\blacken\path(0840,2910)(1200,3000)(0840,3090)(0840,2910)
\path(600,2100)(1200,2100)
\blacken\path(0840,2010)(1200,2100)(0840,2190)(0840,2010)
\path(600,1200)(1200,1200)
\blacken\path(0840,1110)(1200,1200)(0840,1290)(0840,1110)
\path(1425,3000)(2025,3000)
\blacken\path(1665,2910)(2025,3000)(1665,3090)(1665,2910)
\path(3600,3000)(3000,3000)
\blacken\path(3360,3090)(3000,3000)(3360,2910)(3360,3090)
\path(4575,3000)(3975,3000)
\blacken\path(4335,3090)(3975,3000)(4335,2910)(4335,3090)
\path(5475,3000)(4875,3000)
\blacken\path(5235,3090)(4875,3000)(5235,2910)(5235,3090)
\path(5400,2100)(4800,2100)
\blacken\path(5160,2190)(4800,2100)(5160,2010)(5160,2190)
\path(5475,1200)(4875,1200)
\blacken\path(5235,1290)(4875,1200)(5235,1110)(5235,1290)
\path(4800,1800)(4800,1200)
\blacken\path(4710,1560)(4800,1200)(4890,1560)(4710,1560)
\path(1500,1200)(2100,1200)
\blacken\path(1740,1110)(2100,1200)(1740,1290)(1740,1110)
\path(4575,1200)(3975,1200)
\blacken\path(4335,1290)(3975,1200)(4335,1110)(4335,1290)
\path(1200,0975)(1200,0375)
\blacken\path(1110,0735)(1200,0375)(1290,0735)(1110,0735)
\path(2100,0900)(2100,0300)
\blacken\path(2010,0660)(2100,0300)(2190,0660)(2010,0660)
\path(3900,900)(3900,0300)
\blacken\path(3810,0660)(3900,0300)(3990,0660)(3810,0660)
\path(4800,900)(4800,0300)
\blacken\path(4710,0660)(4800,0300)(4890,0660)(4710,0660)
\path(525,4800)(1125,4800)
\blacken\path(0765,4710)(1125,4800)(0765,4890)(0765,4710)
\path(4200,2100)(4800,2100)
\blacken\path(4440,2010)(4800,2100)(4440,2190)(4440,2010)
\path(4800,2400)(4800,3000)
\blacken\path(4890,2640)(4800,3000)(4710,2640)(4890,2640)
\path(4800,3375)(4800,3975)
\blacken\path(4890,3615)(4800,3975)(4710,3615)(4890,3615)
\path(4800,4200)(4800,4800)
\blacken\path(4890,4440)(4800,4800)(4710,4440)(4890,4440)
\path(4500,3900)(3900,3900)
\blacken\path(4260,3990)(3900,3900)(4260,3810)(4260,3990)
\path(3000,5700)(3000,0300)
\path(3900,4200)(3900,4800)
\blacken\path(3990,4440)(3900,4800)(3810,4440)(3990,4440)
\path(3900,5700)(3900,0300)
\path(3300,3900)(3900,3900)
\blacken\path(3540,3810)(3900,3900)(3540,3990)(3540,3810)
\path(3900,3600)(3900,3000)
\blacken\path(3810,3360)(3900,3000)(3990,3360)(3810,3360)
\path(3900,2700)(3900,2100)
\blacken\path(3810,2460)(3900,2100)(3990,2460)(3810,2460)
\path(3000,5100)(3000,5700)
\blacken\path(3090,5340)(3000,5700)(2910,5340)(3090,5340)
\path(3000,4500)(3000,3900)
\blacken\path(2910,4260)(3000,3900)(3090,4260)(2910,4260)
\path(3000,3300)(3000,3900)
\blacken\path(3090,3540)(3000,3900)(2910,3540)(3090,3540)
\path(3000,2400)(3000,3000)
\blacken\path(3090,2640)(3000,3000)(2910,2640)(3090,2640)
\path(3000,900)(3000,300)
\blacken\path(2910,0660)(3000,0300)(3090,0660)(2910,0660)
\path(2100,2700)(2100,2100)
\blacken\path(2010,2460)(2100,2100)(2190,2460)(2010,2460)
\path(2100,3375)(2100,3975)
\blacken\path(2190,3615)(2100,3975)(2010,3615)(2190,3615)
\path(2100,4200)(2100,4800)
\blacken\path(2190,4440)(2100,4800)(2010,4440)(2190,4440)
\path(1200,4275)(1200,4875)
\blacken\path(1290,4515)(1200,4875)(1110,4515)(1290,4515)
\path(1200,3600)(1200,3000)
\blacken\path(1110,3360)(1200,3000)(1290,3360)(1110,3360)
\path(1200,2700)(1200,2100)
\blacken\path(1110,2460)(1200,2100)(1290,2460)(1110,2460)
\path(1200,1800)(1200,1200)
\blacken\path(1110,1560)(1200,1200)(1290,1560)(1110,1560)
\path(3300,2100)(3900,2100)
\blacken\path(3540,2010)(3900,2100)(3540,2190)(3540,2010)
\path(3900,1800)(3900,1200)
\blacken\path(3810,1560)(3900,1200)(3990,1560)(3810,1560)
\path(3600,1200)(3000,1200)
\blacken\path(3360,1290)(3000,1200)(3360,1110)(3360,1290)
\path(0300,3000)(5700,3000)
\path(2400,4800)(3000,4800)
\blacken\path(2640,4710)(3000,4800)(2640,4890)(2640,4710)
\path(2625,3900)(2025,3900)
\blacken\path(2385,3990)(2025,3900)(2385,3810)(2385,3990)
\path(2700,3000)(2100,3000)
\blacken\path(2460,3090)(2100,3000)(2460,2910)(2460,3090)
\path(3000,1500)(3000,2100)
\blacken\path(3090,1740)(3000,2100)(2910,1740)(3090,1740)
\path(2400,2100)(3000,2100)
\blacken\path(2640,2010)(3000,2100)(2640,2190)(2640,2010)
\path(2400,1200)(3000,1200)
\blacken\path(2640,1110)(3000,1200)(2640,1290)(2640,1110)
\path(1500,2100)(2100,2100)
\blacken\path(1740,2010)(2100,2100)(1740,2190)(1740,2010)
\path(2100,1800)(2100,1200)
\blacken\path(2010,1560)(2100,1200)(2190,1560)(2010,1560)
\put(-200,4800){$1$}
\put(-200,3900){$2$}
\put(-200,1200){$N$}
\put(4700,-300){$N$}
\put(2000,-300){$2$}
\put(1100,-300){$1$}
\end{picture}
\begin{ca}
\label{dwbcfigure}
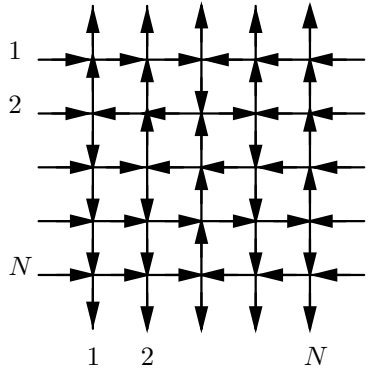
A lattice configuration with domain wall boundary conditions.
\end{ca}
\end{minipage}
\end{center}
\bigskip

\subsection{Korepin's conditions}

In \cite{korepin}, Korepin obtained four conditions that uniquely 
determine $Z_N({\bf x}, {\bf y}, \mu)$. These are

\begin{korepin}
$Z_{N}$ is a symmetric function in the $\{ {\bf x} \}$ and in the 
$\{ {\bf y} \}$ rapidity variables.
\end{korepin}

\begin{korepin}
$Z_{N}$ is a trigonometric polynomial of degree $(N-1)$
in any rapidity variable.
\end{korepin}

\begin{korepin}
Setting $x_1 = y_1 + \mu$, $Z_{N}$ satisfies the recursion relation 
\begin{multline}
\nonumber
Z_{N}\vert_{x_1 = y_1+\mu} =
\\
\ll \pl_{i=2}^{N} \sinh(-x_i + y_1      ) \rr
\sinh \mu
\ll \pl_{j=2}^{N} \sinh(-x_1 + y_j  ) \rr
\\
Z_{(N - 1)} \ll {\hat{x}_1}, {\hat{y}_1}, \mu \rr
\end{multline}

\noindent where ${\hat{x}_1}, {\hat{y}_1}$ indicate
that $x_1$ and $y_1$ are missing from the sets
$\{ {\bf x}\}$ and $\{ {\bf y}\}$.

\end{korepin}

\begin{korepin}
$Z_{1}$ $=$ $\sinh \mu$
\end{korepin}

We refer to \cite{korepin, korepin-book} for details and 
proofs.

\section{$Z_N$ is (a restriction of) a KP $\tau$ function}
\label{determinant}

\subsection{Izergin's determinant expression} Working in
terms of the variables 
$u_i = e^{2 x_i}$, $v_i = e^{2 y_i}$, and $q = e^{-2 \mu}$, 
Izergin's determinant expression is

\begin{equation}
Z_{N}\!\!\!\ll {\bf u}, {\bf v}, q\rr
=
c_N
\frac{
\prod^N_{i,j=1}(u_i-v_j)(q u_i - v_j)
}
{
\prod_{1 \le i < j \le N} (u_i - u_j)(v_j-v_i)}
\textrm{ det}
\left[
\frac{1}{(u_i-v_j)(q u_i-v_j)}
\right]^N_{i,j=1}
\label{izergin-expression}
\end{equation}

\noindent where 

\begin{equation}
c_N = (q-1)^N \prod_{i=1}^{N} u_i^{\frac{1}{2}} v_i^{\frac{1}{2}}
\label{c-n}
\end{equation}

The virtue of Izergin's expression is that it is straightforward 
to show that it satisfies Korepin's conditions 
\cite{izergin, korepin-book}. 

\subsection{Lascoux's determinant expression} 
In \cite{lascoux}, Lascoux showed that Izergin's expression 
for $Z_N$ can be rewritten as 

\begin{equation}
Z_{N}\!\!\!\ll {\bf u}, {\bf v}, q\rr
=
c_N
\textrm{det}
\left[
\sum_{j=1}^{2N-1}
\alpha_{i, j}({\bf u})
\beta_{j, k}({\bf v}, q)
\right]^N_{i, k=1}
\label{lascoux-expression}
\end{equation}

\noindent where

\begin{equation}
\alpha_{i, j}({\bf u}) = h_{-i+j} ({\bf u}),
\quad
\beta_{j, k}({\bf v}, q)
=
\frac{q^{j-k+1}-q^{k-1}}{q-1} 
e_{-j+k+N-1}({\bf - v})
\label{alpha-beta}
\end{equation}

\noindent and $h_n({\bf u})$ and $e_n({\bf v})$ are the 
$n$-th homogeneous and the $n$-th elementary symmetric 
functions, respectively \cite{macdonald}. For a proof 
that Lascoux's expression 
is equivalent to Izergin's, see \cite{lascoux}. As we will 
see shortly, the virtue of Lascoux's expression is that it 
is straightforward to show that it is (a restriction of) 
a KP $\tau$ function.

\subsection{Remarks on notation} In $\alpha_{i, j}$, 
$\beta_{i, j}$, and other matrix elements, $i$ is the 
row and $j$ is the column index. We will omit writing 
the dependence on the parameter $q$ explicitly, as it 
is always there. 

\subsection{Partitions}
We denote the partition of an integer $|\lambda|$ into 
parts $\lambda_1 \geq \lambda_2 \geq \cdots$, 
by 
$[\lambda] = [\lambda_1, \lambda_2, \cdots, \lambda_L]$, 
where $L$ is the number of non-zero parts.
$m^n$ in $[\cdots, m^n, $$\cdots]$ indicates that 
$[\lambda]$ has $n$ parts of length $m$. Further, 
$m^0$ indicates that there are no parts of length $m$ and 
$0^n$ stands for $n$ parts of length $0$.

\subsection{Young diagrams}
Consider the Young diagram $\lambda$ of a partition 
$[\lambda]$ as in Figure {\bf \ref{young-diagram}}. 
The cells are labelled from top to bottom and 
from left to right. The cell with coordinates 
$(i, j),\ i, j \geq 1$ is $c_{ij}$. The number of cells
$c_{11}, c_{22}, \cdots$, on the main diagonal is $d$.
%
%FIG-04
%
\begin{center}
\begin{minipage}{4.8in}
\setlength{\unitlength}{0.001cm}
\renewcommand{\dashlinestretch}{30}
\begin{picture}(4800, 3000)(-4000, 1000)
%
% negative x shifts the figure to the right
% negative y shifts the figure up
\thicklines
%
%The horizontal lines
\path(0000,3600)(2400,3600)
\path(0000,3000)(2400,3000)
\path(0000,2400)(1800,2400)
\path(0000,1800)(1800,1800)
\path(0000,1200)(1200,1200)
%
%The vertical lines
\path(0000,3600)(0000,1200)
\path(0600,3600)(0600,1200)
\path(1200,3600)(1200,1200)
\path(1800,3600)(1800,1800)
\path(2400,3600)(2400,3000)
\put(0050, 3150){\tiny{(1,1)}}
\put(0650, 3150){\tiny{(1,2)}}
\put(1250, 3150){\tiny{(1,3)}}
\put(1850, 3150){\tiny{(1,4)}}
\put(0050, 2550){\tiny{(2,1)}}
\put(0650, 2550){\tiny{(2,2)}}
\put(1250, 2550){\tiny{(2,3)}}
\put(0050, 1950){\tiny{(3,1)}}
\put(0650, 1950){\tiny{(3,2)}}
\put(1250, 1950){\tiny{(3,3)}}
\put(0050, 1350){\tiny{(4,1)}}
\put(0650, 1350){\tiny{(4,2)}}
\end{picture}
\begin{ca}
\label{young-diagram}
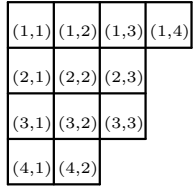
The Young diagram $\lambda = [4, 3, 3, 2] \equiv [ 4, 3^2, 2]$.
The labels $(i, j)$ in a cell $c_{ij}$ are its coordinates. 
In this example, $d=3$.
\end{ca}
\end{minipage}
\end{center}
\bigskip

\subsection{Hooks} Given a cell $c_{ij}$, we define a hook $h_{ij}$ 
as the union of $c_{ij}$, the set of all cells below it and all cells 
to the right of it. We say that the hook $h_{ij}$ has a corner at 
$c_{ij}$.
The diagram on the left hand side of 
Figure {\bf \ref{young-diagram-rectangular}} has 6 hooks, one 
for each cell. The right hand side of 
Figure {\bf \ref{young-diagram-rectangular}} is the hook 
corresponding to the cell $c_{(2, 1)}$.

\subsection{Horizontal $a$-parts and vertical $b$-parts} 
We decompose a hook $h_{ij}$ into a horizontal $a$-part $a_{ij}$ 
of length $|a_{ij}|$ that consists of all cells to the right of 
$c_{ij}$, and a vertical $b$-part $b_{ij}$ of length $|b_{ij}|$ 
that consists of $c_{ij}$ and all cells below it. Hence the allowed 
$a$-part lengths are $0, 1, 2, \cdots$, while the allowed
$b$-part lengths are 
$1, 2, \cdots$ The length of $h_{ij}$ will be indicated
by $|h_{ij}|$ as well, so that
\begin{equation}
|h_{ij}| = |a_{ij}| + |b_{ij}|
\end{equation}

\subsection{Frobenius coordinates of $\lambda$} Decomposing 
each hook $h_{ii}$, $i=1, \cdots, d$ that has a corner at a 
cell on the main diagonal of $\lambda$ into the corner cell 
$c_{ii}$, the set of $m_i$ cells to the right of $c_{ii}$, 
and $n_i$ cells below it, we obtain $2d$ integers 
$\{m_i, n_j\}$ which define the Frobenius coordinates of 
$\lambda$.

%FIG-05
%
\begin{center}
\begin{minipage}{4.8in}
\setlength{\unitlength}{0.001cm}
\renewcommand{\dashlinestretch}{30}
\begin{picture}(4800, 2400)(-3000, 0)
%
% negative x shifts the figure to the right -->
% negative y shifts the figure up
\thicklines
%
%The figure on the left
%The horizontal lines
\path(0000,0000)(1200,0000)
\path(0000,0600)(1200,0600)
\path(0000,1200)(1200,1200)
\path(0000,1800)(1200,1800)
%
%The vertical lines
\path(0000,1800)(0000,0000)
\path(0600,1800)(0600,0000)
\path(1200,1800)(1200,0000)
\put(0050, 1450){\tiny{[1,3]}} \put(0650, 1450){\tiny{[0,3]}}
\put(0050, 0850){\tiny{[1,2]}} \put(0650, 0850){\tiny{[0,2]}}
\put(0050, 0250){\tiny{[1,1]}} \put(0650, 0250){\tiny{[0,1]}}
%
%The diagram on the right -->
%The horizontal lines
%
\path(4000,1200)(5200,1200)
\path(4000,0600)(5200,0600)
\path(4000,0000)(4600,0000)
%
%The vertical lines
\path(4000,1200)(4000,0000)
\path(4600,1200)(4600,0000)
\path(5200,1200)(5200,0600)
%
%\put(4050, 0850){\tiny{[1,2]}} \put(4650, 0850){\tiny{[0,2]}}
%\put(4050, 0250){\tiny{[1,1]}} 
%
\end{picture}
\begin{ca}
\label{young-diagram-rectangular}
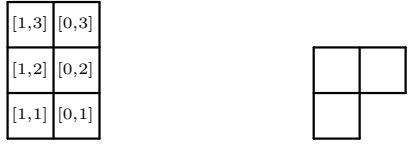
On the left, the Young diagram corresponding to 
$\lambda = [2, 2, 2]$ $ \equiv [2^3]$. 
The labels $[a, b]$ in a cell are the 
$a$-part  and $b$-part of the hook 
associated with that cell. 
On the right, the hook corresponding to 
$c_{21}$ in $\lambda$. In this hook,
the $a$-part $a_{21}$ has length 1, 
the $b$-part $b_{21}$ has length 2.
\end{ca}
\end{minipage}
\end{center}
\bigskip

\subsection{Expanding $Z_N$ in terms of Schur functions}
To make contact with KP theory and eventually obtain an 
expression for $Z_N$ as a vacuum expectation value of 
free fermions, we start by expanding Lascoux's expression 
in terms of Schur functions. Using (\ref{alpha-beta}) 
and the Cauchy-Binet identity in 
(\ref{lascoux-expression}), we obtain

\begin{equation}
Z_{N}\!\!\!\ll {\bf u}, {\bf v}\rr
=
c_N
\sum_{
1 \le j_1 < \cdots< j_{N} \le 2N-1
}
\textrm{det}
\left[
h_{-i + j_l}
({\bf u})
\right]^N_{i, l=1}
\textrm{det}
\left[
\beta_{j_l, k}({\bf v})
\right]^N_{l, k=1}
\label{cauchy-binet}
\end{equation}

Rewriting $j_l$ as $j_l = \lambda_{(N+1) - l} + l$, then  
reversing the order of all rows and all columns in the 
first matrix and transposing it, the sum on the right hand 
side of (\ref{cauchy-binet}) becomes

\begin{equation}
\sum_{
{0 \le \lambda_{N} \leq \cdots}{\leq \lambda_1 \le N-1}
}
\textrm{det}
\left[
h_{\lambda_{i} - i + l}({\bf u})
\right]^N_{i, l=1}
\textrm{det}
\left[
\beta_{\lambda_{(N+1) - l} + l,  k}({\bf v})
\right]^N_{l, k=1}
\end{equation}

Finally, using the definition of Schur functions

\begin{equation}
S_{\lambda}({\bf u})
=
\textrm{det} 
\left[ 
h_{\lambda_i -i + l}({\bf u})
\right]^N_{i, l=1}
\end{equation}

\noindent and (\ref{alpha-beta}) to write

\begin{eqnarray}
c^{(N)}_{\lambda}({\bf v})
&=&
\textrm{det}
\left[
\beta_{\ll \lambda_{l_{\textit{conj}}}\rr - l_{\textit{conj}} + (N+1), k}
({\bf v})
\right]^N_{l, k=1}
\nonumber
\\
&=&
\textrm{det}
\left[ 
\frac{
   q^{ \ll \lambda_{l_{\textit{conj}}}\rr - l_{\textit{conj}} - k + N + 2} 
 - q^{k-1}
     }{q-1}
     \, \, \, 
   e_{-\ll \lambda_{l_{\textit{conj}}}\rr + l_{\textit{conj}} + k
   -2}({\bf - v})
\right]^N_{l, k=1}
\label{c}
\end{eqnarray}

\noindent where 

\begin{equation}
l_{\textit{conj}} = (N+1) - l
\end{equation}

\noindent we end up with 

\begin{equation}
Z_{N}\!\!\!\ll {\bf u}, {\bf v}\rr
=
c_N
\sum_{\lambda \subseteq [(N-1)^N]}
c^{(N)}_{\lambda}({\bf v}) S_{\lambda}({\bf u})
\label{expansion-1}
\end{equation}

Since the summation in (\ref{expansion-1}) is over 
partitions of at most $N$ parts, and there are 
$N$ independent $u$-variables, and 
$N$ independent $v$-variables, there are no terms in the expansion 
that trivially vanish due to lack of sufficiently many independent 
variables. 

\subsection{Expanding $Z_N$ in terms of character polynomials}
Next, following \cite{blue-book}, we introduce the power sum 
variables

\begin{equation}
t_{n} = \frac{1}{n} \sum_{i=1}^{N} u_i^n, \quad n=1, 2, \cdots
\label{power-sums}
\end{equation}

\noindent and the polynomials $p_n ({\bf t})$

\begin{equation}
p_n ({\bf t}) = 
\sum_{n_1 + 2n_2 + 3n_3 + \cdots = n}
\frac{
t_{1}^{n_1} t_{2}^{n_2} t_{3}^{n_3} \cdots
}
{
n_{1}! n_{2}! n_{3}! \cdots
}
\end{equation}

\noindent where $p_i ({\bf t})=0$ for $i<0$. Next, we define 
the polynomials $h_{m, n}({\bf t})$

\begin{equation}
h_{m, n}({\bf t}) := 
(-1)^n 
\sum_{k = 0}^{n} 
p_{k+m+1} ({\bf t}) p_{n-k} ({\bf -t})
\label{hook-polynomials}
\end{equation}

\noindent and finally the character polynomials 
$\chi_{\lambda}({\bf t})$

\begin{equation}
\chi_{\lambda}({\bf t}) 
:= 
\textrm{det} 
\left[
h_{m_i, n_j} ({\bf t})
\right]_{i, j=1}^{d}
\label{character-polynomials}
\end{equation}

\noindent where $\{m_i, n_j\}$ are the Frobenius coordinates
of $\lambda$, and the upper limit $d$ is the number of cells 
on the main diagonal. In terms of character polynomials and 
the $t$-variables of (\ref{power-sums}) (and by 
slightly abusing notation), we rewrite 
$Z_{N}({\bf u}, {\bf v})$ as

\begin{equation}
Z_{N}\!\!\!\ll {\bf t}, {\bf v}\rr
=
c_N
\sum_{\lambda \subseteq (N-1)^N}
c^{(N)}_{\lambda}({\bf v}) \chi_{\lambda}({\bf t})
\label{expansion-3}
\end{equation}

\noindent where $c^{(N)}_{\lambda}({\bf v})$ was defined
in (\ref{c}). We conclude that $Z_N$ 
can be expanded in terms of character 
polynomials, with coefficients that are minors of 
a determinant such that (by construction) they satisfy 
Pl\"ucker relations \cite{blue-book}. If the
$t$-variables were all independent, then 
$Z_N({\bf t}, {\bf v})$ would be identically a KP $\tau$ 
function. We refer to \cite{blue-book} for details 
and proofs.

\subsection{Restricted $t$-variables}
$Z_N({\bf u}, {\bf v})$ is a function of $N$ independent 
$u$-vari-
ables. On the other hand, from the definition of the character 
polynomials, it easy to show that $Z_N({\bf t}, {\bf v})$ 
requires in general more than $N$ $t$-variables. Since 
the latter are power sums of the $N$ $u$-variables, they 
cannot be all independent. Let us write 
$Z_N^{\textit{rest}} ({\bf t}, {\bf v})$ to emphasise 
the fact that the $t$-variables are restricted to be power
sums of a (generally) smaller number of $u$-variables. This 
restriction spoils the interpretation 
of the $t$-variables as KP time variables, and of 
$Z_N^{\textit{rest}}({\bf t}, {\bf v})$
as a KP $\tau$ function.

\subsection{Free $t$-variables}
We can formally consider 
$Z_N^{\textit{rest}}({\bf t}, {\bf v})$ as our starting 
point, take the $t$-variables to be independent, write it as 
$Z_N^{\textit{free}}({\bf t}, {\bf v})$ and think of 
$Z_N^{\textit{rest}}({\bf t}, {\bf v})$ as a restriction of 
$Z_N^{\textit{free}}({\bf t}, {\bf v})$ obtained
by setting the $t$-variables to power sums of the $N$
$u$-variables using (\ref{power-sums}). 
Viewed this way, 
$Z_N^{\textit{free}}({\bf t}, {\bf v})$ 
is a KP $\tau$ function, and 
$Z_N^{\textit{rest}}({\bf t}, {\bf v})$ $=$  
$Z_N({\bf u}, {\bf v})$,
is a restriction of a KP $\tau$ function. 

\section{Fermionic expression for $Z_N$}
\label{ferm}

Given that any KP $\tau$ function can written as the vacuum 
expectation value of exponentials in neutral bilinears in KP 
charged fermions\footnote{By this, we mean linear combinations
of terms of the form $\psi_m \psis_n$.}
\cite{blue-book}, we wish to write $Z_N^{\textit{free}}$ 
as such. Our plan is to propose a vacuum expectation value 
$F_N^{\textit{free}}({\bf t}, {\bf v})$, where we use 
the superscript {\textit{free}} to emphasise that the 
$t$-variables are independent, that by construction is 
a KP $\tau$ function, then show that 
$F_N^{\textit{free}}({\bf t}, {\bf v})$ 
$=Z_N^{\textit{free}}({\bf t}, {\bf v})$.
Hence restricting the $t$-variables in 
$F_N^{\textit{free}}({\bf t}, {\bf v})$
to be powers sums of $N$ $u$-variables, one recovers 
$Z_N^{\textit{rest}}({\bf t}, {\bf v})$, and
$Z_N({\bf u}, {\bf v})$,
as an expectation value of charged free fermions.

\subsection{Charged free fermions}

Consider the free fermion operators $\{\psi_{n}, \psis_{n}\}$, 
$n \in \Z$, with charges $\{+1, -1\}$ 
and energies $n$. They generate a Clifford algebra over 
$\C$ defined by the anti-commutation relations 
\begin{equation}
\left.
\begin{array}{l}
\left[  \psi_m,  \psi_n \right]_{+} = 0 \\ \\ 
\left[ \psis_m, \psis_n \right]_{+} = 0 \\ \\ 
\left[  \psi_m, \psis_n \right]_{+}= \delta_{m,n}
\end{array}
\right\}\ \forall \ m, n \in \Z
\label{aa}
\end{equation}

\subsection{Vacuum states}

We define the vacuum states $\langle 0|$ and $|0\rangle$ by 
the actions

\begin{equation}
\left.
\begin{array}{l}
\langle 0| \psi_{n} = \psi_m |0\rangle = 0, 
\\ \\
\langle 0|\psis_{m} = \psis_n|0\rangle = 0,
\end{array}
\right\} \quad \forall\ m < 0,\ n \geq 0
\label{vacuum-states}
\end{equation}

\noindent and we adopt the inner product normalization
\begin{equation}
\langle 0 | 0 \rangle = 1
\label{norm}
\end{equation}

\subsection{Creation and annihilation operators} 
We refer to the operators that annihilate the vacuum 
state $|0\rangle$,
$\psi_m |0\rangle = 0$, $m = -1, -2,    \ldots$, and
$\psis_n|0\rangle = 0$, $n =  0,  1, 2, \ldots$ 
as {\it annihilation operators}, 
and to the rest,
$\psi_m |0\rangle \neq 0$, $m =  0,  1,  2, \ldots$, and
$\psis_n|0\rangle \neq 0$, $n = -1, -2,     \ldots$
as {\it creation operators}. 

\subsection{Normal ordering}
The normal-ordered product is defined, as usual, by placing 
annihilation operators to the right of creation operators
\begin{equation*}
:\psi_{i}\psis_{j}:\ = \psi_{i}\psis_{j}-\langle 0|
\psi_{i}\psis_{j}|0\rangle
\end{equation*}

\subsection{The Heisenberg algebra} Consider the operators 
$H_m$, 

\begin{equation}
H_m := {\sum_{j \in \Z}:\psi_{j}\psis_{j+m}:}, 
\quad m \in \Z 
\label{ak}
\end{equation}

\noindent that together with the central element $1$ form 
a Heisenberg algebra 

\begin{equation}
\left[ H_m, H_n \right] = m \delta_{m + n, 0}, 
\ \forall \ m, n \in \Z 
\label{al}
\end{equation}

\subsection{The Hamiltonian}
Using $H_m, m = 1, 2, 3, \dots$, we define the Hamiltonian

\begin{equation}
H(\mathbf{t}) := \sum_{m = 1}^{\infty} t_m H_m
\end{equation}

\subsection{Fermion bilinears} Given a partition 
$\lambda$, we can associate to each hook $h_{ij}$ a bilinear 
$\psi_{|a_{ij}|} \psis_{-|b_{ij}|}$ with net charge zero. 
Both fermions in the bilinear are creation operators, 

\begin{equation}
\psi_{|a_{ij}|} \psis_{-|b_{ij}|}
=
- \psis_{-|b_{ij}|} \psi_{|a_{ij}|}
\end{equation}

\noindent and normal ordering is unnecessary. 

\subsection{Boson-fermion correspondence}
It is possible to realise fermionic Fock space 
expressions as elements in the polynomial ring 
$\field{C} [t_1, t_2, \cdots]$ \cite{blue-book}. 
For the purposes of this work, all we need is that fact that
the character polynomial $\chi_{\lambda}({\bf t})$ can be
generated as follows

\begin{equation}
\langle 0| 
e^{H({\bf t})} 
\psis_{-b_1}  \cdots  \psis_{-b_d}
 \psi_{ a_d}  \cdots   \psi_{ a_1} 
|0\rangle 
= 
(-1)^{b_1 + \cdots + b_d} \chi_{\lambda}({\bf t}) 
\label{boson-fermion}
\end{equation}

\noindent where 
$a_d < \cdots < a_1$ and 
$b_d < \cdots < b_1$ 
are the $a$-part and $b$-part lengths of the hooks associated 
with cells $c_{11}, c_{22}, \cdots$, on the main diagonal of 
$\lambda$, and $d$ is the number of these cells. In terms of 
Frobenius coordinates, 
$a_i = m_i$ and 
$b_j = n_j + 1$.

\subsection{Partitions and fermion monomials}
Consider the fermion monomial consisting of 
$d$ $\psi$ and 
$d$ $\psis$ fermions, 
$\psis_{-b_1}  \cdots  \psis_{-b_d}
  \psi_{ a_d}  \cdots   \psi_{ a_1}$, where
  $a_d < \cdots < a_1$ and
  $b_d < \cdots < b_1$
as on the left hand side of (\ref{boson-fermion}).
The fermions in the monomial are ordered according to 
length and correspond to the $a$-parts 
and $b$-parts of the partition $\lambda$ on the right hand 
side of (\ref{boson-fermion}), with the longer 
$b$-parts to the left and the longer $a$-parts to the right.
We refer to such a monomial associated to a partition 
$\lambda$ as a partition-ordered, or $\lambda$-ordered
monomial. The partition in Figure {\bf \ref{young-diagram}}
has $d=3$, 
a set of $b$-parts $\{4, 3, 1\}$ and 
a set of $a$-parts $\{3, 1, 0\}$. 
The corresponding monomial is
$\psis_{-4} \psis_{-3} \psis_{-1}
  \psi_{ 0}  \psi_{ 1}  \psi_{ 3}
$.

\subsection{A fermion expectation value}
Consider the vacuum expectation value 
\begin{equation}
F_N^{\textit{free}} ({\bf t}, {\bf v}) = 
\langle0|
e^{H({\bf t})}
e^{X^{(N)}_{0}}
e^{X^{(N)}_{1}}
\cdots
e^{X^{(N)}_{N-2}}
|0\rangle
\label{fermion-expectation-value}
\end{equation}
\noindent where
\begin{equation}
X^{(N)}_{a}
=
\sum_{b=1}^{N}
(-)^b 
d^{(N)}_{[a+1, 1^{b-1}]} 
\psis_{-b} \psi_a,
\quad
a = 0, 1, \cdots, N-2
\label{X}
\end{equation}
\noindent where $1^0$ inside a partition $[\lambda]$ 
stands for a trivial, length-0 part, and

\begin{equation}
d^{(N)}_{\lambda} = c^{(N)}_{\lambda}/c^{(N)}_{\phi}
\label{d}
\end{equation}

The coefficient $c^{(N)}_{\lambda}$ was defined in (\ref{c}), 
and the (infinitely many) $t$-parameters in
({\ref{fermion-expectation-value}}) are taken as 
independent variables rather than power sums of the 
(finitely many) $u$-parameters. 
Since $F_N^{\textit{free}}$ is an expectation value of 
exponentials of neutral bilinears in $\{\psi_n, \psis_{m}\}$, 
which are generators of $gl(\infty)$, followed by the time 
evolution operator $e^{H({\bf t})}$, it is by construction 
a KP $\tau$ function \cite{blue-book}. 
Since all fermions in the linear combinations of bilinears 
in (\ref{X}) are creation operators, there is no need for 
normal ordering these terms.

\subsection{A combinatorial interpretation}
The exponentials that appear in $F_N^{\textit{free}}$ in 
(\ref{fermion-expectation-value}) have the 
following interpretation.
Consider a rectangular partition of $N$ rows and $(N-1)$ 
columns. For example, for $N=3$, see the diagram on the
left hand side of Figure {\bf \ref{young-diagram-rectangular}}.
For the $j$-th column, with $N$ cells of $a$-part length 
$a =$ $(N-1)-j$,
we have an exponential 
$X^{(N)}_a$, $0 \le a \le (N-2)$, 
which is a sum of $N$ terms. The $b$-th term is of the form 
$(-1)^{b} d^{(N)}_{[h]} \psis_{-b} \psi_{a}$,
where $[h]$ is the partition corresponding to the hook 
that has a corner at that cell, and $a$ and $b$ are the 
$a$-part and $b$-part lengths of that hook.

\subsection{$F_N$ is proportional to $Z_N$}
\label{lemma}
In the rest of this work, we show that

\begin{equation}
c_N
c^{(N)}_{\phi}      ({\bf v}) 
F_N^{\textit{free}} ({\bf t}, {\bf v})  = 
Z_N^{\textit{free}} ({\bf t}, {\bf v})
\label{main}
\end{equation}

\subsection{The normalization factor} 
Expanding the exponentials in (\ref{fermion-expectation-value}) 
and using $\langle 0| 0\rangle$ $= 1$, the character polynomial 
expansion of $F_N^{\textit{free}}$ starts with 1.  On the other 
hand, since $\chi_{\phi} ({\bf t})=1$, the character polynomial 
expansion of $Z_N$ in (\ref{expansion-3}) starts with the 
$t$-independent term, $c_N c_{\phi}^{(N)}({\bf v})$. 
We choose to keep the definition of the vertex weights as in 
(\ref{weights}), and show that $F_N$ and $Z_N$ are equal up 
to a factor of $c_N c^{(N)}_{\phi}({\bf v})$.

\subsection{Example: $N=3$}
Before we prove (\ref{main}) for general $N$, 
we verify it in the simple but non-trivial case of $N=3$. 
Omitting the superscripts $(3)$ to simplify the notation, 
we have

\begin{eqnarray}
c^{}_{\lambda}
&=&
\textrm{det}
\left[
\frac{q^{ \ll \lambda_{4-i} \rr  + i - j + 1} - q^{j-1}}{q-1} 
     e_{- \ll \lambda_{4-i} \rr  - i + j + 2}({\bf - v}) 
\right]^3_{i, j=1}
\\
X^{}_{0} 
&=& 
-d^{}_{[1,0,0]} \psis_{-1} \psi_0
+d^{}_{[1,1,0]} \psis_{-2} \psi_0 
-d^{}_{[1,1,1]} \psis_{-3} \psi_0 
\nonumber
\\
X^{}_{1} 
&=& 
-d^{}_{[2,0,0]} \psis_{-1} \psi_1
+d^{}_{[2,1,0]} \psis_{-2} \psi_1
-d^{}_{[2,1,1]} \psis_{-3} \psi_1
\nonumber
\end{eqnarray}

\noindent where $d^{}_{\lambda}$ was defined in (\ref{d}). 
Notice that in this case, the relevant partition is shown on the 
left hand side of Figure {\bf \ref{young-diagram-rectangular}}, 
and the bilinears in $X^{}_0$ and $X^{}_1$ correspond to the hooks 
of the cells in the two columns. This leads to

\begin{eqnarray}
\label{complicated}
e^{X^{}_0}
e^{X^{}_1}
=
1
&-& d^{}_{[1,0,0]} \psis_{-1} \psi_0
 +  d^{}_{[1,1,0]} \psis_{-2} \psi_0
 -  d^{}_{[1,1,1]} \psis_{-3} \psi_0
\\
&-& d^{}_{[2,0,0]} \psis_{-1} \psi_1
 +  d^{}_{[2,1,0]} \psis_{-2} \psi_1
 -  d^{}_{[2,1,1]} \psis_{-3} \psi_1
\nonumber
\\
&+&
\ll
 d^{}_{[1,1,0]} d^{}_{[2,0,0]}
-d^{}_{[1,0,0]} d^{}_{[2,1,0]}
\rr
\psis_{-2} \psis_{-1} \psi_0 \psi_1
\nonumber
\\
&+&
\ll
 d^{}_{[1,0,0]} d^{}_{[2,1,1]}
-d^{}_{[1,1,1]} d^{}_{[2,0,0]} 
\rr
\psis_{-3} \psis_{-1} \psi_0 \psi_1
\nonumber
\\
&+&
\ll
 d^{}_{[1,1,1]} d^{}_{[2,1,0]}
-d^{}_{[1,1,0]} d^{}_{[2,1,1]}
\rr
\psis_{-3} \psis_{-2} \psi_0 \psi_1
\nonumber
\end{eqnarray}

Consider  the six fermion bilinears on the right hand side 
of (\ref{complicated}). Using the boson-fermion 
correspondence, each of these bilinears corresponds 
to a partition $[\lambda]$ (which in the case of 
bilinears is a single hook), and its coefficient 
$d_{[\lambda]}$ corresponds to the same partition 
$[\lambda]$. If the same is true for the three fermion 
quartic terms, then from (\ref{expansion-3}), 
$F_3$ would be proportional to $Z_3$. To put the 
coefficients of the quartic terms in the right form, 
we need to show that

\begin{eqnarray}
 d^{}_{[1,1,0]} d^{}_{[2,0,0]}
-d^{}_{[1,0,0]} d^{}_{[2,1,0]}
&=& 
-               d^{}_{[2,2,0]}
\label{p1}
\\
 d^{}_{[1,0,0]} d^{}_{[2,1,1]}
-d^{}_{[1,1,1]} d^{}_{[2,0,0]}
&=&
\phantom{-}
                d^{}_{[2,2,1]}
\label{p2}
\\
 d^{}_{[1,1,1]} d^{}_{[2,1,0]}
-d^{}_{[1,1,0]} d^{}_{[2,1,1]}
&=&
-               d^{}_{[2,2,2]}
\label{p3}
\end{eqnarray}

\subsection{Pl\"ucker relations}
Equations (\ref{p1}--\ref{p3}) are examples of Pl\"ucker 
relations, which are identities involving bilinears in 
$n$$\times$$n$ minors of a $2n$$\times$$2n$ matrix with 
zero determinant. In this case, $n=3$. For details, see 
\cite{hirota}. Using the notation 

\begin{equation}
c^{}_{\lambda}  
\equiv
c^{}_{[\lambda_1, \lambda_2, \lambda_3]}
=
\textrm{det}
\left[ 
\gamma_{\lambda_3 +1}, \gamma_{\lambda_2 +2}, \gamma_{\lambda_1 +3}
\right]
=
\left| 
\gamma_{\lambda_3 +1}, \gamma_{\lambda_2 +2}, \gamma_{\lambda_1 +3}
\right|,
\label{notation}
\end{equation}

\noindent where $\gamma_{b}$ is the 3-component column vector

\begin{equation}
\gamma_{b} =
\ll
\frac{
q^{b-a+1} - q^{a-1}
}
{
q-1
} 
e_{-b+a+2}({\bf -v})
\rr_{a= 1, 2, 3},
\end{equation}

\noindent we consider the $6\times 6$ determinant

\begin{equation}
\left| \begin{array}{cccccc}
\gamma_{\mu_1} &
\gamma_{\mu_2} &
\gamma_{\nu_1} &
\gamma_{\nu_2} &
\gamma_{\nu_3} &
\gamma_{\nu_4} \\
0 &
0 &
\gamma_{\nu_1} &
\gamma_{\nu_2} &
\gamma_{\nu_3} &
\gamma_{\nu_4}
\end{array} \right| = 0,
\end{equation}

\noindent and Laplace expand it as the sum of bilinears 
in $3\times3$ determinants

\begin{eqnarray}
\label{generator}
\left| \gamma_{\mu_1}, \gamma_{\mu_2}, \gamma_{\nu_1} \right|
\left| \gamma_{\nu_2}, \gamma_{\nu_3}, \gamma_{\nu_4} \right| 
-
\left| \gamma_{\mu_1}, \gamma_{\mu_2}, \gamma_{\nu_2} \right|
\left| \gamma_{\nu_1}, \gamma_{\nu_3}, \gamma_{\nu_4} \right| 
&&
\\
+
\left| \gamma_{\mu_1}, \gamma_{\mu_2}, \gamma_{\nu_3} \right|
\left| \gamma_{\nu_1}, \gamma_{\nu_2}, \gamma_{\nu_4} \right| 
-
\left| \gamma_{\mu_1}, \gamma_{\mu_2}, \gamma_{\nu_4} \right|
\left| \gamma_{\nu_1}, \gamma_{\nu_2}, \gamma_{\nu_3} \right| 
&=&
0
\nonumber
\end{eqnarray}

We can use (\ref{generator}) to generate the
Pl\"ucker relations that we need by suitably choosing 
the column vectors $\gamma_b$. 

\subsection{The $N=3$ Pl\"ucker relations}
\label{generating}
The required Pl\"ucker relation, defined as in 
(\ref{first}), has a term $d_{[\lambda]} d_{[\phi]}$, where 
$d_{[\phi]} =1$, and 
$\lambda = [2, 2, 0]$. From (\ref{notation}) 
we see that the bilinear that we need is 
$|\gamma_{1+0}, \gamma_{2+2}, \gamma_{3+2}|
 |\gamma_{1+0}, \gamma_{2+0}, \gamma_{3+0}|$ 
 $=$
$|\gamma_{1  }, \gamma_{4  }, \gamma_{5  }|
 |\gamma_{1  }, \gamma_{2  }, \gamma_{3  }|$. 
This dictates the choice of parameters 

\begin{equation}
(\mu_1, \mu_2, \nu_1, \nu_2, \nu_3, \nu_4)
= 
(    1,     4,     5,     1,     2,     3) 
\label{choice-1}
\end{equation}

Using (\ref{choice-1}) in (\ref{generator}), 
and the antisymmetry property of determinants, we obtain 

\begin{equation}
\left| \gamma_{1}, \gamma_{4}, \gamma_{5} \right| 
\left| \gamma_{1}, \gamma_{2}, \gamma_{3} \right|
-
\left| \gamma_{1}, \gamma_{3}, \gamma_{5} \right| 
\left| \gamma_{1}, \gamma_{2}, \gamma_{4} \right|
=
-
\left| \gamma_{1}, \gamma_{3}, \gamma_{4} \right| 
\left| \gamma_{1}, \gamma_{2}, \gamma_{5} \right|
\label{first}
\end{equation}

\noindent which, suing (\ref{d}) and (\ref{notation}), 
is the Pl\"ucker relation that we need to prove (\ref{p1}). 

Equations (\ref{p2}--\ref{p3}) are obtained by setting
$(\mu_1, \mu_2, \nu_1, $$\nu_2, $$\nu_3, $$\nu_4)$ to
$(2, 4, 5, 1, 2, 3)$, then to
$(3, 4, 5, 1, 2, 3)$ in (\ref{generator}). 
Finally, using (\ref{boson-fermion}), we 
obtain

\begin{equation}
c^{}_{\phi}
\langle 0|
e^{H({\bf t})}
e^{X^{}_{0}}
e^{X^{}_{1}}
|0\rangle
=
\sum_{\lambda \subseteq [2,2,2]}
c^{}_{\lambda} ({\bf v}) 
\chi_{\lambda}({\bf t})
\label{ee}
\end{equation}

The right hand side of (\ref{ee}) is Lascoux's form expanded 
as in (\ref{expansion-3}).
The above example shows that the proof of (\ref{main}) 
relies on using Pl\"ucker relations to simplify the coefficients 
of expansions of the exponentials that appear in 
(\ref{fermion-expectation-value}). 

\subsection{Plan of proof of Equation (\ref{main})}
\label{main-proof}
Firstly, we expand the product of two exponentials in terms of 
neutral monomials in the fermions, with coefficients that can 
be simplified using Pl\"ucker relations, then we use induction 
to show that this can be done for the product of $N$ exponentials. 
Finally we use boson-fermion correspondence to obtain Lascoux's 
form of $Z_N$.

\subsection{Remark on notation} In the following, we consider 
only the general $N$ case, so we omit the superscripts $(N)$, 
and write $X_a$, instead of $X_a^{(N)}$, and so on.  

\subsection{Expanding two exponentials}

For $0 \le a_2 < a_1 \le (N-2)$, consider the product

\begin{equation}
e^{X^{}_{a_2}} e^{X^{}_{a_1}} 
=
1 + X^{}_{a_2} + X^{}_{a_1} + X^{}_{a_2} X^{}_{a_1}
\label{two}
\end{equation}

\noindent where 

\begin{eqnarray}
X^{}_{a_1} 
&=&
\sum^{N}_{b=1}(-1)^{b}
d^{}_{[a_1+1,1^{(b-1)}]} \psis_{-b} \psi_{a_1}
\\
X^{}_{a_2} 
&=&
\sum^{N}_{b=1}(-1)^{b}
d^{}_{[a_2+1,1^{(b-1)}]} \psis_{-b} \psi_{a_2}
\nonumber
\\
X^{}_{a_2} X^{}_{a_1} 
&=&
\sum^N_{
\ttop{b_1, b_2 = 1}{b_1 \ne b_2}
}
(-1)^{b_1+b_2} 
d^{}_{[a_2+1,1^{(b_2-1)}]} d^{}_{[a_1+1,1^{(b_1-1)}]}
\psis_{-b_2} \psi_{a_2} \psis_{-b_1} \psi_{a_1}
\nonumber
\end{eqnarray}

Note that while we know that $a_2 < a_1$, we can have 
$b_2 < b_1$ and $b_1 < b_2$. We account for these two
possibilities by rewriting the cross-term as 

\begin{multline}
\label{two-terms-1}
X^{}_{a_2} X^{}_{a_1}
=
\\
\ll 
\sum_{1 \le b_2 < b_1 \le N} + \sum_{1 \le b_1 < b_2 \le N}
\rr
(-1)^{b_1 + b_2} 
d^{}_{
[a_2+1, 1^{(b_2-1)}]
} 
d^{}_{
[a_1+1, 1^{(b_1-1)}]
}
\\
\times
\psis_{-b_1} \psis_{-b_2} \psi_{a_2}  \psi_{a_1}
\end{multline}

Relabelling the indices in the second summation, so 
that $b_1 \rightarrow b_2$ and $b_2 \rightarrow b_1$,
and using the anticommutation relations to put the 
product of fermion operators in the form used in 
(\ref{boson-fermion}), the right hand side 
of (\ref{two-terms-1}) becomes

\begin{multline}
\label{two-terms-2}
\sum_{1 \le b_2 < b_1 \le N}
(-1)^{b_1+b_2}
\\
\ll
d^{}_{
[a_2+1, 1^{(b_2-1)}]
}
d^{}_{
[a_1+1, 1^{(b_1-1)}]
} 
-
d^{}_{
[a_2+1, 1^{(b_1-1)}]
}
d^{}_{
[a_1+1, 1^{(b_2-1)}]
} 
\rr
\\
\times
\psis_{-b_1} \psis_{-b_2} 
 \psi_{ a_2}  \psi_{ a_1}
\end{multline}

\noindent we see that once again we need to simplify 
the coefficients. Following the same ideas that were
used in the $N=3$ example above, we write

\begin{equation}
c^{}_{\lambda}  =
\left|
\gamma_{\ll \lambda_{N  } \rr +1},
\gamma_{\ll \lambda_{N-1} \rr +2},
\cdots,
\gamma_{\ll \lambda_{1  } \rr +N}
\right|
\label{5.20}
\end{equation}

\noindent where $\gamma_{b}$ is the $N$-component 
column vector
\begin{equation}
\gamma_{b} =
\ll
\frac{
q^{b-a+1}-q^{a-1}
}
{
q-1
} 
e_{-b+a+N-1} ({\bf -v}) 
\rr_{a = 1, \cdots, N}
\label{5.19}
\end{equation}

\noindent then consider the $2N$$\times$$2N$ determinant 
expression

\begin{equation}
\left| 
\begin{array}{ccccccc}
\gamma_{\mu_1} & \cdots & \gamma_{\mu_{N-1}} & \gamma_{\nu_1} &
\gamma_{\nu_2} & \cdots & \gamma_{\nu_{N+1}}
\\
 0             & \cdots &  0                 & \gamma_{\nu_1} & 
\gamma_{\nu_2} & \cdots & \gamma_{\nu_{N+1}}
\end{array} 
\right| 
= 0
\end{equation}

\noindent and Laplace expand it as a sum of bilinears in 
$N$$\times$$N$ determinants 

\begin{equation}
\sum^{N+1}_{p=1} 
(-1)^{p+1} 
\left| 
\gamma_{\mu_1} \cdots  \cdots \gamma_{\mu_{N-1}} \gamma_{\nu_p}
\right|  
\left| 
\gamma_{\nu_1} \cdots \hat{\gamma}_{\nu_{p}} \cdots \gamma_{\nu_{N+1}}
\right| 
= 0
\label{general-plucker}
\end{equation}

\noindent where $\hat{\gamma}$ indicates a missing term. 

\subsection{The general $N$ Pl\"ucker relation}
The procedure outlined for the $N=3$ case in subsection 
{\bf \ref{generating}}, can be extended to general $N$. 
From (\ref{two-terms-2}), the Pl\"ucker relation 
that we need contains a bilinear term 
$d_{[\lambda]} d_{[\phi]}$, where $\lambda$ has 
two $a$-parts, $a_1 > a_2$ and 
two $b$-parts, $b_1 > b_2$, so that 

\begin{equation}
\lambda = [a_1 + 1, 
           a_2 + 2, 
           2^{b_2 - 2}, 
           1^{b_1 - b_2 -1},
           0^{N - b_1}],
\end{equation}

\noindent where we have indicated the 0-length elements as 
          well to make the total length of the partition 
	  $N$. We prepare a $2N$-element sequence $\Sigma^0$

\begin{equation}
\Sigma^0 
=
[1, 2, \ldots, N | 1, 2, \ldots, N] 
\label{initial-sequence}
\end{equation}

\noindent where the vertical bar $|$ in (\ref{initial-sequence})
emphasizes the structure of $\Sigma^{0}$ as a concatenation of 
a left  sequence $\Sigma^0_{\textit{left }}$ $=$ $[1, 2, \ldots, N]$ 
and 
a right sequence $\Sigma^0_{\textit{right}}$ $=$ $[1, 2, \ldots, N]$. 
Given $\lambda = [\lambda_1, \cdots, \lambda_N]$,
where $\lambda_1 \ge \lambda_2 \ge \cdots \ge \lambda_N$, 
we modify $\Sigma^{0}_{\textit{left }}$ as follows. 
We add 
$\lambda_1$ to the largest integer ($N$) in 
$\Sigma^0_{\textit{left }}$, 
$\lambda_2$ to the next largest ($N-1$), and so forth. 
$\Sigma^0_{\textit{right}}$ remains unmodified.
This generates a new sequence $\Sigma^{}$. 
Choosing the parameters in the $2N$-element sequence
$(\mu_1, \cdots, \mu_{N-1}, \nu_1, $ $\cdots $ $\nu_{N+1})$
by identifying them with the corresponding integers in
$\Sigma^{}$, we end up with

\begin{eqnarray}
\label{parameters-2}
(\mu_1, \cdots, \mu_{N-b_1})           
&=& 
(1, \cdots, N-b_1)
\\
(\mu_{N-b_1+1}, \cdots, \mu_{N-b_2-1}) 
&=& 
(N-b_1 +2, \cdots, N-b_2)
\nonumber
\\
(\mu_{N-b_2}, \cdots, \mu_{N-2})       
&=& 
(N-b_2 +2, \cdots, N)
\nonumber
\\
(\mu_{N-1}, \nu_{1})                   
&=& 
(N+a_2 +1, N+a_1+1)
\nonumber
\\
(\nu_2, \cdots, \nu_{N+1})             
&=& (1, \cdots, N)
\nonumber
\end{eqnarray}

\noindent where \lq $ \cdots $\rq\ between two integers
$m$ and $n$ stands for the integers $m+1,$$ m+2,$$ 
\cdots,$$ n-1$, that increase by 1 at a time.
Using (\ref{general-plucker}), the above choice 
of parameters leads to the Pl\"ucker relation

\begin{multline}
\left| 
\Sigma,
\gamma_{N+a_1 +1}
\right|
\left|
\gamma_1, 
\cdots, 
\gamma_N 
\right|
\\
+ (-1)^{N-b_1} 
\left| 
\Sigma,
\gamma_{N-b_1 +1} 
\right|
\left| 
\gamma_{N+a_1 +1}, 
\gamma_1, 
\cdots, 
\hat{\gamma}_{N-b_1 +1}, 
\cdots, 
\gamma_N 
\right|
\\
+ (-1)^{N-b_2} 
\left| 
\Sigma,
\gamma_{N-b_2 +1} 
\right|
\left| 
\gamma_{N+a_1 +1}, 
\gamma_1,
\cdots, 
\hat{\gamma}_{N-b_2 +1}, 
\cdots, 
\gamma_N 
\right| 
\\
= 0
\label{something}
\end{multline}

\noindent where $\Sigma$ in (\ref{something}) is 
the sequence 

\begin{equation}
\Sigma = 
\{
\gamma_{1      }, \cdots, \gamma_{N-b_1},
\gamma_{N-b_1+2}, \cdots, \gamma_{N-b_2},
\gamma_{N-b_2+2}, \cdots, \gamma_{N    },
\gamma_{N+a_2+1}
\}
\end{equation}

Using (\ref{notation}), (\ref{something}) can 
be written as 

\begin{multline}
c^{}_{[a_2+1,1^{b_2-1}]} c^{}_{[a_1+1,1^{b_1-1}]} -
c^{}_{[a_2+1,1^{b_1-1}]} c^{}_{[a_1+1,1^{b_2-1}]}
=  
\\
c^{}_{\phi}
c^{}_{[a_1+1, a_2+2, 2^{b_2-1}, 1^{b_1-b_2-1}]}
\label{plucker-2}
\end{multline}

Substituting (\ref{plucker-2}) in the second line 
of (\ref{two-terms-1}), we obtain 

\begin{equation}
X^{}_{a_2}  X^{}_{a_1} =  
\sum_{1 \le b_2 < b_1 \le N}(-1)^{b_2+b_1}
d^{}_{[a_1+1,a_2+2,2^{b_2-1},1^{b_1-b_2-1}]}
\psis_{-b_1} \psis_{-b_2} \psi_{a_2} \psi_{a_1}
\label{p.22}
\end{equation}

\subsection{Expanding more than two exponentials}
We wish to show that for all 
$0 \le a_k < \cdots < a_1 \le N-2$,
we have 

\begin{equation}
X^{}_{a_k} \cdots X^{}_{a_1}
=
\sum_{
{1 \le b_k < \cdots}{ < b_1 \le N}
}
(-1)^{b_1 + \cdots + b_{k}}
d^{}_{\lambda_{(d=k)}}
\psis_{-b_1} \cdots \psis_{-b_k} \psi_{a_k} \cdots \psi_{a_1}
\label{proposition}
\end{equation}

\noindent where

\begin{equation}
\lambda_{(d=k)}
=
[a_{1}+1, 
 a_{2}+2,
\cdots,
 a_{k}+k, 
     k^{b_{k  }       - 1}, 
 (k-1)^{b_{k-1} - b_k - 1},
\cdots,
     1^{b_{1  } - b_2 - 1},
     0^{N       - b_1    }
]
\end{equation}

\noindent is the partition, with $k$ cells on the main 
diagonal, associated with the fermion monomial on the 
right hand side of (\ref{proposition}).

We start by assuming that (\ref{proposition}) holds 
for some value of $k$, and show that it also holds for $k+1$. 
Since it holds for $k=2$, it holds for all $k$. 
Consider the expression

\begin{multline}
X^{}_{a_k} \cdots X^{}_{a_1} X^{}_{a_0} =
\sum_{
      {1 \le b_k < \cdots < b_1 \le N}
                                      }
\sum^{N}_{
\ttop{b_0 = 1}{b_{0} \ne b_1, \cdots, b_k}
}
(-1)^{b_0 + b_1 + \cdots + b_k}
\\
d^{}_{\lambda_{(d=k)}}
d^{}_{[a_{0} +1, 1^{b_{0}-1}]}
\psis_{-b_1} 
\cdots 
\psis_{-b_k} 
 \psi_{ a_k} 
\cdots
 \psi_{ a_1}
\psis_{-b_0}
 \psi_{ a_0}
\end{multline}

\noindent for $0 \le a_k < \cdots < a_1 < a_0 \le (N-2)$,
where $\lambda_{(d=k)}$ is still the same partition as in 
(\ref{proposition}), and rewrite the summation as

\begin{multline}
\sum_{
{1 \le b_k < \cdots < b_1 \le N}
} 
\sum^N_{
\ttop{b_{0} = 1}{b_{0} \ne b_1, \cdots, b_k}
} 
= 
\\
\sum_{
{1 \le b_k < \cdots < b_0 \le N}
} +  
\sum_{
{1 \le b_k < \cdots < b_2 < b_0 < b_1 \le N}
} + 
\cdots + 
\\
\sum_{
{1 \le b_k < b_0 < b_{k-1} < \cdots < b_1 \le N}
} + 
\sum_{
{1 \le b_0 < b_k < \cdots < b_1 \le N}
}
\end{multline}

Ordering the fermions in each summation, so that they are 
in the same order as in (\ref{boson-fermion}), 
and relabeling the indices, we obtain

\begin{multline}
\sum_{
{1 \le b_k < \cdots < b_0 \le N}
}
(-1)^{b_0 + \cdots + b_k}
\ll
\sum^k_{p=0}
(-1)^{p}
d^{}_{\lambda_{(d=k|p)}}
d^{}_{[
a_0 + 1, 1^{(b_p-1)}
]}
\rr
\\
\times
\psis_{-b_0}
\cdots
\psis_{-b_k}
 \psi_{ a_k}
\cdots
 \psi_{ a_0}
\label{complicated-expression}
\end{multline}

\noindent where $\lambda_{(d=k|p)}$
stands for $\lambda_{(d=k)}$, but the $b$-part that used 
to be $b_p$ is now omitted and the $b$-part $b_0$ is 
added in correct position. For example, if $\lambda_{(d=6)}$ 
has $b$-parts $\{b_6 < b_5 < b_4 < b_3 < b_2 < b_1\}$,
then
$\lambda_{(d=6|3)}$ has 
$\{b_6 < b_5 < b_4 < b_2 < b_1 < b_0\}$.
$\lambda_{(d=k|0)}$ indicates that no changes were made.

To simplify the sum in large brackets in 
(\ref{complicated-expression}), we consider 
(\ref{general-plucker}), and, following the same procedure 
as before, use the evaluations 

\begin{eqnarray}
(\mu_1, \cdots, \mu_{N - b_0})
&=&
(1, \cdots, N-b_0)
\label{sequences}
\\
(\mu_{N - b_0 + 1},\cdots, \mu_{N - b_1 -1})
&=&
(N - b_0 + 2, \cdots, N - b_1)
\nonumber
\\
(\mu_{N - b_1}, \cdots, \mu_{N - b_2 - 2})
&=&
(N - b_1 + 2, \cdots, N - b_2)
\nonumber
\\
(\mu_{N - b_2 - 1}, \cdots, \mu_{N - b_ 3- 3})
&=&
(N - b_2 + 2, \cdots,  N - b_3)
\nonumber
\\
& \vdots&
\nonumber
\\
(\mu_{N - b_k - (k-1)}, \cdots, \mu_{N - (k+1)})
&=&
(N - b_k + 2, \cdots, N)
\nonumber
\\
(\mu_{N-k}, \cdots,  \mu_{N-1}, \nu_1)
&=&
(N + a_k + 1, \cdots, N + a_1 +1, N + a_0 +1)
\nonumber
\\
(\nu_2, \cdots, \nu_{N+1})
&=&
(1, \cdots, N)
\nonumber
\end{eqnarray}

The required Pl\"ucker relation is 

\begin{eqnarray}
& &
\left| \Sigma, \gamma_{N+a_0+1} \right|
\left| \gamma_1,  \cdots, \gamma_N                          \right| 
\label{big-plucker}
\\
&+&
(-1)^{N - b_0 + 1}   
\left| \Sigma, \gamma_{N-b_0+1} \right|
\left| \gamma_{N+a_0+1}, \cdots, \hat{\gamma}_{N-b_0+1}, \cdots, 
                                           \gamma_N         \right|
\nonumber
\\
&+&
(-1)^{N -b_1 + 1}  
\left| \Sigma, \gamma_{N-b_1+1} \right|
\left| \gamma_{N+a_0+1}, \cdots, \hat{\gamma}_{N-b_1+1}, \cdots, 
                                           \gamma_N         \right|
\nonumber
\\
&+&
\cdots 
\nonumber
\\
&+&
(-1)^{N - b_k + 1} 
\left| \Sigma, \gamma_{N-b_k+1} \right| 
\left| \gamma_{N+a_0+1}, \cdots, \hat{\gamma}_{N-b_k+1}, \cdots, 
                                           \gamma_N         \right|  
\nonumber
\\
&=& 
0
\nonumber
\end{eqnarray}

\noindent where the sequence $\Sigma$ in (\ref{big-plucker}) 
is obtained as follows. Consider the sets of indices 
in (\ref{sequences}) from top to bottom except the 
very last, $(1, \cdots, N)$, and order them from 
left to right, so that the top-most 
$(1, \cdots, N-b_0)$ 
becomes the left-most and 
$(N+a_k+1, \cdots, N+a_1+1, N+a_0+1)$ becomes 
the right-most. Concatenate them and remove 
the right-most element $(N+a_0+1)$. Use the
resulting $(N-1)$-element sequence in the same
order to generate the indices of $\Sigma$ 
in (\ref{big-plucker}). Using (\ref{notation}), 
(\ref{big-plucker}) can be rewritten as 

\begin{equation}
\sum^k_{p=0}(-1)^{p} 
c^{}_{\lambda_{(d=k|p)}}
c^{}_{[a_0 + 1, 1^{b_p - 1}]} =  
c^{}_{\phi} c^{}_{ \lambda_{(d= k+1)} }
\end{equation}

\noindent where $\lambda_{(d=k+1)}$ is the partition 
with $(k+1)$ cells on the main diagonal

\begin{multline}
\lambda_{(d=k+1)} =
[
a_0 + 1, a_1 + 2, 
\cdots, 
a_k +(k+1), 
(k+1)^{b_k    -1}, 
 k^{   b_{k-1} - b_k - 1}, 
 \cdots, 
1^{b_0 - b_1 -1} 
]
\end{multline}

Relabeling $a_i$ and $b_j$ so that the indices 
take values from 1 to $(k+1)$, rather than from 
0 to $k$, such that $a_1 > a_2 > \dots > a_{k+1}$ 
and $b_1 > b_2 > \dots > b_{k+1}$, we obtain

\begin{equation}
X^{}_{a_{k+1}} \cdots X^{}_{a_1} =
\sum_{
1 \le b_{k+1} < \cdots < b_1 \le N
}
(-1)^{b_1+ \cdots + b_{k+1}} 
d^{}_{
       \lambda_{(d=k+1)}
                         } 
\psis_{-b_1}     \cdots \psis_{-b_{k+1}} 
 \psi_{ a_{k+1}} \cdots  \psi_{ a_1}
\end{equation}

\noindent where $\lambda_{(d=k+1)}$ now stands for the partition
that corresponds to the partition-ordered monomial adjacent to it. 
This completes the proof of (\ref{proposition}). 

\subsection{Expanding $(N-1)$ exponentials}
Using (\ref{proposition}), we obtain 

\begin{multline}
\label{p.26}
e^{X^{}_{0}} \cdots e^{ X^{}_{N-2}} =
1 +
\sum^{N-1}_{d=1}
\sum_{
{0 \le a_d < \cdots}{ < a_1 \le N-2}
}
X^{}_{a_d} \cdots X^{}_{a_1}
\\
=
1 + 
\sum^{N-1}_{d=1} 
\sum_{
\ttop{0 \le a_d < \cdots < a_1 \le (N-2)}
     {1 \le b_d < \cdots < b_1 \le  N   }
}
(-1)^{b_1 + \cdots + b_{d}} 
d^{}_{\lambda_{d}} 
\psis_{-b_1} \cdots \psis_{-b_d}
 \psi_{ a_d} \cdots  \psi_{ a_1}
\end{multline}

From the remarks following (\ref{boson-fermion}), 
we note that the terms on the right hand side of 
(\ref{p.26}) generate every fermionic expression that 
corresponds to any partition that fits in the rectangle
$[(N-1)^{N}]$. 
Additionally, (\ref{proposition}) shows that the 
fermionic expressions are accompanied by the required 
coefficient with the required sign. Thus

\begin{multline}
\langle 0|
e^{H({\bf t})} e^{X^{}_0}
\cdots
e^{X^{}_{N-2}} 
|0 \rangle 
\\
=
d^{}_{\phi}({\bf v}) \chi_{\phi} ({\bf t}) +
\sum^{N-1}_{k=1} 
\sum_{
{0 \le a_k < \cdots}{ < a_1 \le N-2}
}
\sum_{
{1 \le b_k < \cdots}{ < b_1 \le N}
}
d^{}_{\lambda_{k}}({\bf v}) \chi_{\lambda_{k}}({\bf t})
\\
=
\sum_{\lambda \subseteq (N-1)^{N}}
d^{}_{\lambda}({\bf v})\chi_{\lambda}({\bf t})
\label{final}
\end{multline}

Multiplying both sides of (\ref{final}) 
by $c_N c_{\phi}$, we obtain (\ref{main}). 

\section{Discussion}
\label{discussion}

The result of this note is that a central object in the 
algebraic Bethe Ansatz (ABA) approach to the six vertex 
model (or more precisely the XXZ spin chain), that 
is the domain wall partition function $Z_N$, is 
(a restriction of) a KP $\tau$ function, for all values 
of the crossing parameter. In a sense, what we do is to 
propose a KP $\tau$ function that depends on infinitely 
many time variables, then restrict the time variables to 
be power sums of exponentials of the ABA auxiliary space 
rapidities of $Z_N$ to obtain Lascoux's determinant 
expression of the latter. 

This of course is not the first time that a connection 
between objects in a solved statistical mechanical model
and in a classical integrable hierarchy is observed. For 
pioneering works in this direction, see \cite{korepin-book} 
and references therein. But in earlier studies that involved 
the six vertex model or XXZ spin chain, a connection with 
an integrable nonlinear differential equation (in that case, 
the nonlinear Schr\"odinger equation) was made at the free 
fermion point \cite{korepin-book}. The connection with KP 
that we obtain in this note is valid for all values of the 
crossing parameter.

We also obtained an expression for $Z_N$ as an expectation 
value of exponentials of bilinears in the KP fermions. It 
remains to use this expression to obtain new results, but 
we observe that the bilinears that appear have a simple 
combinatorial interpretation. This leads us to believe that 
analogous expressions can be found for other interesting 
objects in the ABA approach to the six vertex model and 
XXZ spin chain such as the scalar products and norms 
of Bethe vectors. 

\section*{Acknowledgements}
We thank Professors M~Jimbo, N~Kitanine and P~Zinn-Justin for 
useful remarks on the subject of this work and related topics. 
MW and MZ are supported by Australian Postgraduate Awards and 
the Department of Mathematics and Statistics,  The University 
of Melbourne.

\end{document}